\documentclass[11pt,aps,showpacs,showkeys,preprintnumbers,amsmath,amssymb,nofootinbib]{revtex4}

\usepackage{subfigure}
\usepackage{graphicx}
\usepackage{epstopdf}
\usepackage{color}
\usepackage{amsmath}
\usepackage{amsfonts}
\usepackage{amssymb}
\def \be  {\begin{equation}}
\def \ee  {\end{equation}}
\def \ee  {\end{equation}}
\def \bea {\begin{eqnarray}}
\def \eea {\end{eqnarray}}

\begin{document}

\vspace*{0.5cm}

\title{Rapidity Distribution within Landau hydrodynamical model and EPOS event-generator at RHIC Energies }

\author{R. M. Abdel Rahman}
\affiliation{Modern Academy for Engineering and Technology, Basic Sciences Department, 11571, Mokattam, Cairo, Egypt}

\author{Abdel Nasser Tawfik}
\email{atawfik@cern.ch}
\affiliation{Nile University - Egyptian Center for Theoretical Physics (ECTP), Juhayna Square off 26th-July-Corridor, 12588 Giza, Egypt}

\author{Mahmoud Y. El-Bakry}
\affiliation{Ain Shams University, Faculty of Education, Physics Department, 11771, Roxy, Cairo, Egypt}

\author{D. M. Habashy}
\affiliation{Ain Shams University, Faculty of Education, Physics Department, 11771, Roxy, Cairo, Egypt}

\author{Mahmoud Hanafy}
\affiliation{Physics Department, Faculty of Science, Benha University, 13518, Benha, Egypt}

\date{\today}

\begin{abstract}  

The rapidity distribution of well-identified particles such as pions, kaons, protons and their antiparticles measured in the BRAHMS experiment (Au+Au collisions), at $\sqrt{s_{NN}}$ = $62.4$ and $200$ GeV, are compared with that obtained from huge statistical ensembles of 100, 000 events generated from the Cosmic Ray MonteCarlo (CRMC) EPOS event-generator. All these data are then compared to the results calculated in the frame work of Landau hydrodynamical model. We conclude that the Landau hydrodynamical model is capable of describing both experimental data and EPOS $1.99$ event-generator results, and they are in a good agreement. 

\end{abstract}


\maketitle


\section{Introduction}

Quantum chromodynamics (QCD) introduces essential hypothesis to the strong interactions between the fundamental particles, i.e., quarks which mediated by gluons, that make up hadrons \cite{SHURYAK198071,Braun-Munzinger:2015hba,Tawfik:2019qrx}. As the distance between partons decreases, the QCD coupling constant between particles weakens, resulting in asymptotic freedom. For sufficiently high energy density (above the nuclear ground state), a new state of partonic matter with relativistic gas properties is likely to be formed, in which the color degrees-of-freedom (dof) are no longer limited and the parton distribution follows Stefan-Boltzman statistics\cite{Andronic:2014zha,Diamantini:2018mjg}, and known as quark-gluon plasma (QGP) \cite{Blaizot:1999bv}. 
 
The large number of generated particles, and their subsequent interactions and correlations, either at the partonic or hadronic level, persuade the use of fluid dynamics in their interpretation in ultra-relativistic heavy-ion collisions. More pions, kaons and protons are created in such heavy-ion nuclear collisions \cite{Bearden:2004yx}. The yields of these light mesons are more effective as indicators of the entropy and strangeness induced in reactions than the yields of baryons \cite{Blume:2004vq}. 

These particles, on the other hand, are regarded as sensitive indicators of the presence of an early colored deconfined phase of QGP.  \cite{Landau:104093}. At hadronization stage, all hadrons are generated directly from QGP \cite{Koch:1986hf,Koch:1986ud,Kuznetsova:2006bh}. In order to avoid violating conservation rules, it is necessary to gauge the majority of the created particles when assessing the source qualities \cite{Bearden:2001xw}. The relative abundances and kinematic characteristics of generated particles provide a valuable tool for determining if equilibrium occurs during the impact \cite{Bearden:2004yx}. 

In high-energy experiments, the secondary particles which are produced are likely distributed along the beam axis (in the $z$-direction) \cite{Koide_2016}. The rapidity distribution is a best way to describing the boosted angular distribution \cite{Bearden:2004yx}. The rapidity variable has the useful property as it transformed linearly under Lorentz transformation \cite{Fu:2018qsk}. There are some quantities that affecting the the rapidity property such as the freezeout chemical potential and the temperature \cite{Fu:2018qsk} while others probably not affecting the rapidity such as the viscosity. Rapidity would also manifest the possible change in the nature of the expanding colliding system.

The rapidity distribution is calculated according to Carruther's approach in ref. \cite{Tawfik:2019qrx}. On the one hand, Fermi's statistical thermal models can neither represent the dynamical evolution of the collision nor cover more observables, which encourage Landau to investigate the hydrodynamical properties of expanding matter formed in heavy-ion collisions in 1953 \cite{HAMA2005}. The Landau hydrodynamical model \cite{Back:2001bq} is one of the most well-known approaches  that describes the properties of a system in a hydrodynamical evolution in nuclear heavy-ion collisions \cite{Bearden:2001xw,Steinberg_2005,Carruthers:1973ws,Wong:2008ex}. Ref. \cite{article1} has analyzed the experimental data of the rapidity distribution measured at two distinct RHIC energies in the frame work of Landau hydrodynamical model.

In the present study, we have supposed that the fireball is thermally stable and expands hydrodynamically until it reaches the pion freeze-out density \cite{Landau:104093}. The Landau hydrodynamical model predicts a good approximated Gaussian shaped rapidity distribution undergoing an isentropic (constant entropy) expansion governed by an equation of state \cite{Blume:2004vq}. This model is divided into two parts: the initial longitudinal expansion (during which transverse displacements and velocities are disregarded) and the subsequent "conic flight" in which transverse velocities appear and hadronic particles freezeout. 

In this work, we have estimated the rapidity distribution of well-identified particles such as pions, kaons, protons and their antiparticles from large statistical ensembles of $100, 000$ events deduced from the Cosmic Ray MonteCarlo (CRMC) EPOS event-generator \cite{PhysRevC.74.044902,Pierog:2013ria} and then compared to that measured from BRAHMS experiment (Au+Au collisions) at $\sqrt{s_{NN}}$ = $62.4$ and $200$ GeV. The aim of this comparison is to conclude that wether the EPOS $1.99$ event-generator can be used to predict the rapidity distribution for particles at energies where the experiment is absent or not. All these data are fitted with the Landau hydrodynamical model.

The following is a breakdown of the paper's structure. In Sec. \ref{models}, We will present our rapidity distribution estimation methods, which are based on the Landau hydrodynamical model (Sec. \ref{sec:LHmodel}) and the EPOS $1.99$ event-generator (Sec. \ref{sec:crmc}). In Sec. \ref{Results and Discussion}, Landau hydrodynamical model results on rapidity distribution for particles, pions, kaons, protons and their antiparticles, produced from Au+Au nuclear heavy-ion collisions are presented in comparison with both BRAHMS experiment data and EPOS $1.99$ event-generator results. The ultimate conclusions are discussed in Sec. \ref{conc}.
\section{Approaches}
\label{models}
Hydrodynamics is regularly utilized to depict the advancement of the collision mechanism, from an initial time (initial condition) to a freeze-out time. Hydrodynamical models are referred to the models that depend on fluid dynamic principle. In this way, hydrodynamical models are used to describe the heavy-ion collisions concerning macroscopic parameters such as temperatures, chemical potentials and flow velocities, related to concepts of equilibrium and expansion.
The used approaches are Landau hydrodynamical model and CRMC-EPOS event-generator model which will be explained in details in the next subsections.
  
\subsection{Landau hydrodynamical model} 
\label{sec:LHmodel}  
The collision is usually depicted in nuclear fluid models as being at local equilibrium and maintaining energy-momentum conservation \cite{article2}. These models are useful for high-density particles, which occur when particle separation is smaller than the mean free path in the medium. The fundamental benefit of the hydrodynamical models is that it has been proved that a huge number of degrees of freedom contained within the microscopic composition of fluids may be reduced to a few macroscopic hydrodynamical variables that characterize the fluid's local attribute such as viscosity \cite{deSouza:2015ena}. 

The equation of state (EoS) in hydrodynamical models describes the relationship between pressure and energy density \cite{Steinberg:2004vy}. 
The EoS may or may not include a phase transition, and it is assumed that the fluid is in or very close to local equilibrium \cite{PhysRevLett.32.741}.
The collision of two strongly Lorentz contracted systems was reported by Fermi and Landau in the 1950s \cite{Landau:104093,Belenkji1956,Carruthers:1973rw}. These beginning circumstances were placed in the context of three-dimensional relativistic hydrodynamics by Landau \cite{Jiang:2013rm}. It is assumed that the evolution is isentropic, and viscosity is not taken into account \cite{Wong:2008ta}. In contrast to Fermi's vision, where there is no hydrodynamic evolution, this description is known as the Landau hydrodynamical model \cite{Landau:104093}.

The fluid elements with the least rapidity magnitude will enter the second stage first, followed by those with the largest rapidity magnitude \cite{Milekhin:1959,Belenkji1956}. The dynamics of the system can be followed up to the endpoint of the hydrodynamical evolution by matching the answers for the first and second phases.
 
A new analysis of the experimental results has proved that Landau hydrodynamical approach is able to produce good conclusions which are consistent with the experimental measurements \cite{Steinberg:2004wx,Murray:2004gh,Jiang:2013rm, Wong:2008ta}. A quantitative analyses of a hydrodynamical system could be made by using an approximate style of the Landau's rapidity distribution. In Landau approach, the equation that describes the relationship between the rapidity and the rapidity distribution is estimated from the experimental measurements which has a Gaussian distribution form.
 
The majority of the experimental data is well represented by Landau's energy dependent Gaussian rapidity distribution, according to a detailed analysis of the data as \cite{Sahoo:2016hln,Landau:104093,Steinberg:2004vy}

\begin{equation}
\frac{1}{\sigma}\frac{d\sigma}{dy}=\frac{dN}{dy}, \label{equ:onee}
\end{equation}
where $\sigma$ represents the width of the rapidity distribution.

By integrating the single-particle inclusive distribution across the transverse momentum $p_\bot$ and dividing by the inelastic cross section, the rapidity distribution of the $i$-th particle $dN^i / dy$  is derived as \cite{Cleymans:2007jj}

\begin{equation}
\frac{dN^i}{dy}=\int_{-\infty}^\infty\rho(y_{FB})\frac{dN^i_1(y-y_{FB})}{dy}dy, \label{equ:twoo}
\end{equation}

where $y_{FB}$, $y$ are the rapidity of fireballs and beam particles, respectively.
$\frac{dN^i_1y}{dy}$ can be obtained from the  distribution of a thermal single fireball.

The distribution of the fireballs that centered at zero, $\rho(y_{FB})$, in a Gaussian distribution form is defined as \cite{Cleymans:2007jj}
\begin{equation}
\rho(y_{FB})=\frac{1}{\sqrt{2\pi}\sigma}\exp(-\frac{y_{FB}^2}{2\sigma^2}),    \label{equ:threee}
\end{equation} 

The Lorentz invariant momentum space volume element is given by \cite{Rafelski:2005hrg}

\begin{equation}
\frac{d^3p}{E}=m_{\bot}dm_{\bot}dyd\phi=p_{\bot}dp_{\bot}dyd\phi,  \label{equ:fourr}
\end{equation}
where $p_{\bot}$ and $m_{\bot}$ are the transverse momentum and transverse mass, respectively. $\phi$ is the azimuthal angle.
The rapidity variable has the advantage of transforming linearly when subjected to a Lorentz transformation where the differential particle momentum distribution is defined as \cite{Rafelski:2005hrg}

\begin{equation}
E\frac{d^3N}{d^3p}=E\frac{d^3\sigma}{p_{\bot}dp_{\bot}dp_{z}d\phi}=\frac{d^3\sigma}{p_{\bot}dp_{\bot}d(y-y_{FB})d\phi}=\frac{d^3\sigma}{m_{\bot}dm_{\bot}d(y-y_{FB})d\phi},    \label{eqU:fivee}
\end{equation}

where 
 \begin{equation}
d(y-y_{FB})=\frac{dp_{\bot}}{E},  \label{equ:sixx}
\end{equation}

and \begin{equation}
p_{\bot}dp_{\bot}=m_{\bot}dm_{\bot},  \label{equ:sivenn}
\end{equation}

In terms of rapidity distribution, Eq. (\ref{eqU:fivee}) can be rewritten as \cite{Rafelski:2005hrg}
\begin{equation}
E\frac{d^3\sigma}{d^3p}=\frac{d^3N}{m_{\bot}dm_{\bot}d(y-y_{FB})d\phi}=\frac{d^2N}{2 \pi m_{\bot}dm_{\bot}d(y-y_{FB})}=\frac{d^2N}{2 \pi p_{\bot}dp_{\bot}d(y-y_{FB})},    \label{eqU:eightt}
\end{equation}

The phase-space distribution is given by \cite{Rafelski:2005hrg}

\begin{equation}
E\frac{d^3\sigma}{d^3p} \equiv f(E, p_z), \label{equ:ninee}
\end{equation}
where $p_z$ means that the produced particle can characterize the axis of collision. The parameter $E$ ensures that the function $f$ is invariant under Lorentz transformation.

Landau discovered a link between beam energy and the number of created charged particles per pair of participants. The rapidity distribution of the created charged particles is written as \cite{PhysRevC.87.044902,Gao:2013dca,Sahoo:2014aca}

\begin{equation}
\frac{dN}{dy}\simeq\exp(\sqrt{L^{2} - y^{2}}),
\label{equ:one}
\end{equation}
where $L$ is the logarithm of Lorentz contract factor $\gamma$ \cite{Landau:104093,Steinberg:2004vy}, and it consider the measure of the thickness of the Lorentz contracted disks of the colliding hadronic matter \cite{Sahoo:2014aca}.
\begin{equation}
L=\ln(\gamma)= \ln(\frac{\sqrt{s_{NN}}}{2m_{p}})= \sigma_{y}^{2},
\label{equ:two}
\end{equation}
where $\sqrt{s_{NN}}$ is the center of mass energy, $\sigma_{y}$ is the width of the distribution, and $m_{p}$ is the mass of proton \cite{Cleymans:2007jj}.

It should be noted that Landau considered pseudorapidity $\eta$ \cite{Landau:104093} and Shuryak first one who introduced rapidity instead \cite{Gao:2013dca}. The rapidity was also introduced by others such as Milekhin who consider $y=\eta$ at high energies \cite{Milekhin:1959}.

Firstly, this relationship is qualitatively consistent with the experimental findings. Lately, for the purpose of quantitative analysis of the hydrodynamical evolution and after a series of calculated treatments, Eq. (\ref{equ:one}) is approximately taken as a Gaussian distribution and expressed as \cite{Landau:104093,Stachel:1989pa,Netrakanti:2005iy,Wong:2008ta,Jiang:2013rm,Gao:2013dca}

\begin{equation}
\frac{dN}{dy}= \frac{N}{\sqrt{2\pi L}}\exp(-\frac{y^{2}}{2L}).
\label{equ:three}
\end{equation}

where $N$ is the normalization constant.

However, this equation is an approximate representation of Eq. (\ref{equ:one}) in the region of $|y|<< L$, these distributions are totally different in other rapidity regions \cite{Jiang:2013rm}.

The values of the fitting parameters, i.e., $N$ and $L$, represented in Eq. (\ref{equ:three}), which are estimated in the frame work of ROOT data analysis \cite{root}, are summarized in Tab. (\ref{tab1}).
\subsection{Cosmic Ray Monte Carlo (CRMC) model}
\label{sec:crmc}

EPOS is a parton model that creates parton ladders through numerous binary parton-parton interactions. Energy-sharing for cross-section calculations, particle creation, parton multiple scattering, outshell remnants, screening and shadowing via unitarization and splitting, and collective effects for dense media are all integrated into EPOS. CRMC (Cosmic Ray Monte Carlo) is a project that includes an interface to cosmic ray models for effective QCD-like models, the Pierre Auger Observatory, and high-energy experiments including NA61, LHCb, TOTEM, ATLAS, and CMS. The cosmic ray models, such as EPOS $1.99$/LHC, are developed on top of the Gribov-Regge model \cite{Drescher:2000ha}. CRMC provides a supplemental description of the backdrop, including diffraction, as well as a common interface for accessing the output from various nuclear collision event generators. The interface is linked to a variety of models, but it is particularly focused on models that use simulations of large cosmic ray air showers, such as qgsjet$01$ \cite{kalmykov1997quark,kalmykov1993nucleus}, qgsjetII \cite{ostapchenko2006nonlinear,ostapchenko2006qgsjet}, sibyII \cite{engel1992nucleus,fletcher1994s,ahn2009cosmic}, EPOS $1.99$ \cite{werner2006parton,pierog2009epos}, QGSJET$01$, and SIBYII$2.3$, at low energies. EPOS $1.99$/LHC and QGSJETII v$03$ and v$04$ are the ones to use at high energies. 

EPOS is a cosmic ray air shower model that may be used to study pp- and AA-collisions at SPS, RHIC, and LHC energies. EPOS employs a simplified treatment of interactions in the final state and can be applied to heavy-ion interactions with least bias hadronic interactions \cite{pierog2015epos}. It's worth noting that EPOS, even in its final form, does not provide simulations for whole hydro systems. The EPOS $1.99$/LHC model, which was used in the calculations, contains a huge number of parameters that describe the fundamental quantities in physics as well as the phenomenological postulates. Experimental and theoretical assumptions can be used to correct these issues. EPOS $1.99$/LHC is considered to build a plausible picture of hadronic interactions based on data supplied by available experiments and event generators.  

In this paper, we utilize the EPOS $1.99$ event generator, in a standard parameter calculations, for two different center-of-mass energies, $62.4$ and $200$ GeV for Au+Au collisions.
We've generated ensembles of at least $100,000$ in events (at each of these energies). We have focused on the effect of collective hadronization in nucleus-nucleus collisions at RHIC in this script. As a result, we'll employ EPOS $1.99$ event-generator in our calculations.

\section{Results and Discussion}
\label{Results and Discussion}  

The current study is based on a replication of the rapidity distribution experiment results of $dN/dy$ of $\pi^{-}$ \cite{BRAHMS:2009acd,Murray:2004gh,Lee:2004bx,Bearden:2004yx}, $\pi^{+}$ \cite{BRAHMS:2009acd,Murray:2004gh,Lee:2004bx,Bearden:2004yx}, $k^{-}$ \cite{BRAHMS:2009acd,Murray:2004gh,Lee:2004bx,Bearden:2004yx}, $k^{+}$ \cite{BRAHMS:2009acd,Murray:2004gh,Lee:2004bx,Bearden:2004yx}, $\bar{p}$ \cite{Arsene:2009aa,Murray:2004gh,Song:2007cx,Bearden:2003hx,PhysRevLett.90.102301}, and $p$ \cite{Arsene:2009aa,Murray:2004gh,Song:2007cx,Bearden:2003hx,PhysRevLett.90.102301} at RHIC energies for Au + Au collisions at $62.4$ GeV and $200$ GeV. We also compare our results to those estimated from the EPOS $1.99$ event-generator \cite{werner2006parton,pierog2009epos}. After that, both sets of results are confronted to the Landau hydodynamical analysis.
The present work aims to study the rapidity distribution in the frame work of Landau hydrodynamical model which has  Gaussian or an exponential form as appear in Eq. (\ref{equ:three}). It is clear that for all particles the dependence of the rapidity distribution $dN/dy$ on the rapidity $y$ is fitted to Gaussian distribution. The values of the fitting parameters, i.e., $N$ and $L$, which shown in Eq. (\ref{equ:three}), that make the Landau hydrodynamical model show a good fit with the experimental measurements at the considered energies for all used particles are summarized in Tab. (\ref{tab1}). The fitting parameters of the Landau hydrodynamical model is estimated in the frame work of ROOT data analysis \cite{root}.

\begin{figure}[htb]
\includegraphics[width=0.6\linewidth]{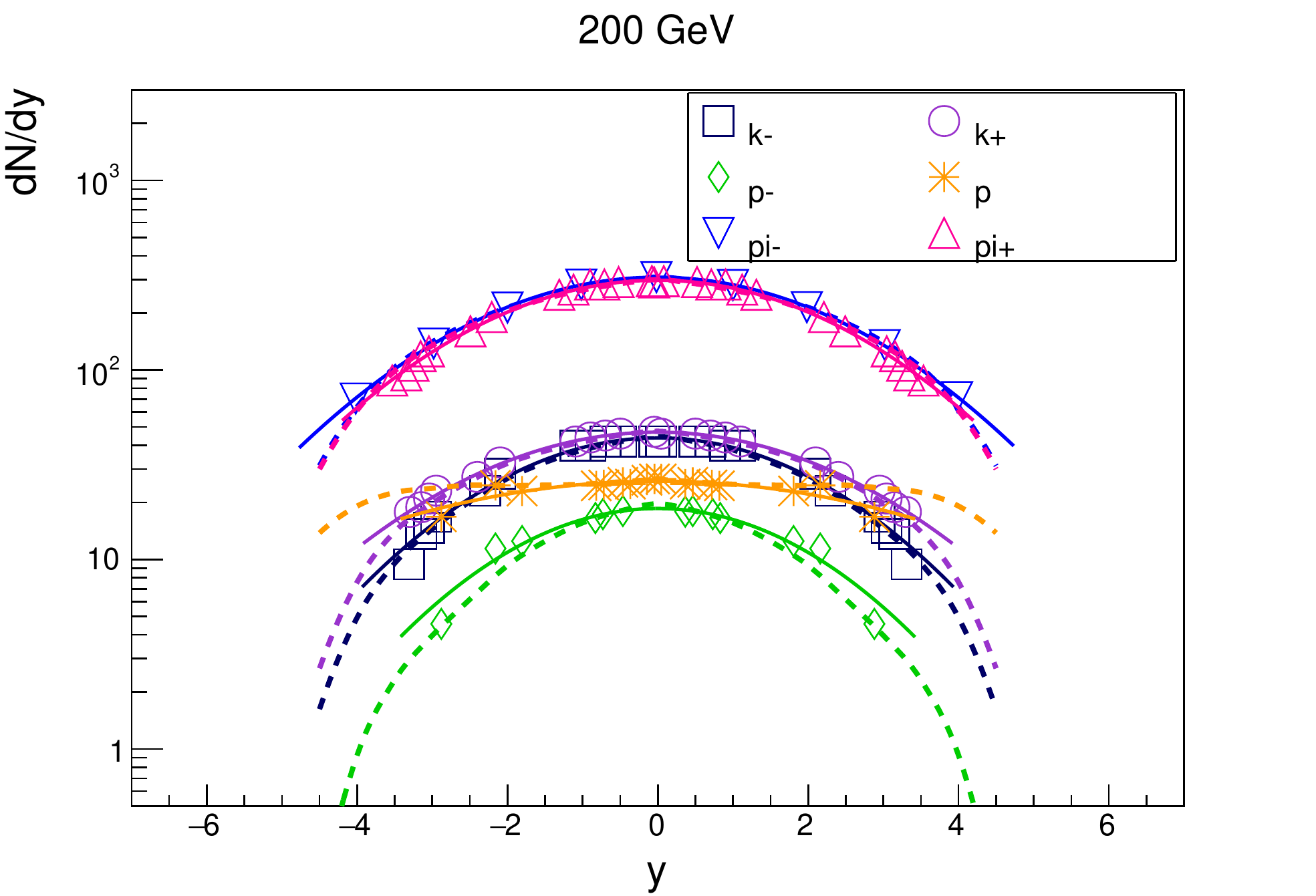}
\caption{The rapidity distribution $dN/dy$ for $\pi^{-}$ \cite{Murray:2004gh,Lee:2004bx,Bearden:2004yx}, $\pi^{+}$ \cite{Murray:2004gh,Lee:2004bx,Bearden:2004yx}, $k^{-}$ \cite{Murray:2004gh,Lee:2004bx,Bearden:2004yx}, $k^{+}$ \cite{Murray:2004gh,Lee:2004bx,Bearden:2004yx}, $\bar{p}$ \cite{Arsene:2009aa,Murray:2004gh,Song:2007cx,Bearden:2003hx,PhysRevLett.90.102301}, and $p$ \cite{Arsene:2009aa,Murray:2004gh,Song:2007cx,Bearden:2003hx,PhysRevLett.90.102301} is depicted as a function of the rapidity. The experimental results (symbols) from Au+Au central collisions, at $200$ GeV, are compared to the Cosmic Ray Monte-carlo (CRMC EPOS $1.99$) event-generator (solid lines), section \ref{sec:crmc}, and also fitted to the Landau hydrodynamical approach (dashed lines), Eq. (\ref{equ:three}).}
\label{fig:one}
\end{figure}

Fig. \ref{fig:one} shows the rapidity distribution $dN/dy$ vs rapidity $y$ for the well-identified hadrons $\pi^{-}$ \cite{Murray:2004gh,Lee:2004bx,Bearden:2004yx}, $\pi^{+}$ \cite{Murray:2004gh,Lee:2004bx,Bearden:2004yx}, $k^{-}$ \cite{Murray:2004gh,Lee:2004bx,Bearden:2004yx}, $k^{+}$ \cite{Murray:2004gh,Lee:2004bx,Bearden:2004yx}, $\bar{p}$ \cite{Arsene:2009aa,Murray:2004gh,Song:2007cx,Bearden:2003hx,PhysRevLett.90.102301}, and $p$ \cite{Murray:2004gh,Song:2007cx,Bearden:2003hx,PhysRevLett.90.102301} measured in Au+Au central collisions from BRAHMS experiment data, at $200$ GeV at rapidity range $0<y<5$, while the left points at range $-5<y<0$ are the reflection points with respect to the mid-rapidity. Experimental data are represented as symbols, while the Landau results are represented as dashed lines and the result from large statistical ensemble of $100000$ events of the EPOS $1.99$ event-generator \cite{werner2006parton,pierog2009epos} are depicted as solid lines.

We observe that the rapidity distribution curves for protons and antiprotons lay at bottom of the graph, while the curves of kaons are above protons curves and below the pions curves, which is positioned at top of the graph. Accordingly, the particles of heavier masses seems to have less rapidity than the particles with lighter masses. On the other hand, for the particles of the same mass but opposite charges, the results show that there are differences between their curves. The particles with heavier masses have higher rapidity distributions. Therefore, the curves of antiprotons lay below that of the protons. The differences decrease for kaons also where the positive kaon curves lay above the negative ones and that difference disappears in case of pions. The proton curves show a different behavior in rapidity distribution which exhibit Gaussian distributed about $y\simeq 3$ apart from the mid-rapidity. It seems that the reactions are showed high degree of transparency causes the formation of a baryon free region at mid-rapidity with approximate balance between matter and anti-matter \cite{Staszel:2005aw}. Landau model results from Eq. ( \ref{equ:three}) shows the same behavior. Finally, the EPOS $1.99$ event-generator \cite{werner2006parton,pierog2009epos} results seem to coincide with experimental data \cite{Murray:2004gh,Lee:2004bx,Bearden:2004yx,Song:2007cx,Bearden:2003hx,PhysRevLett.90.102301} and to the fitted results from Landau hydrodynamical model. Also, the Landau hydrodynamical model shows a good fit with both the experimental measurements and to that estimated from the EPOS $1.99$ event-generator through the obtained minimal chi-square values for the fitting parameters, $N$ and $L$, presented in Tab. (\ref{tab1}).

\begin{figure}[htb]
\includegraphics[width=.6\linewidth]{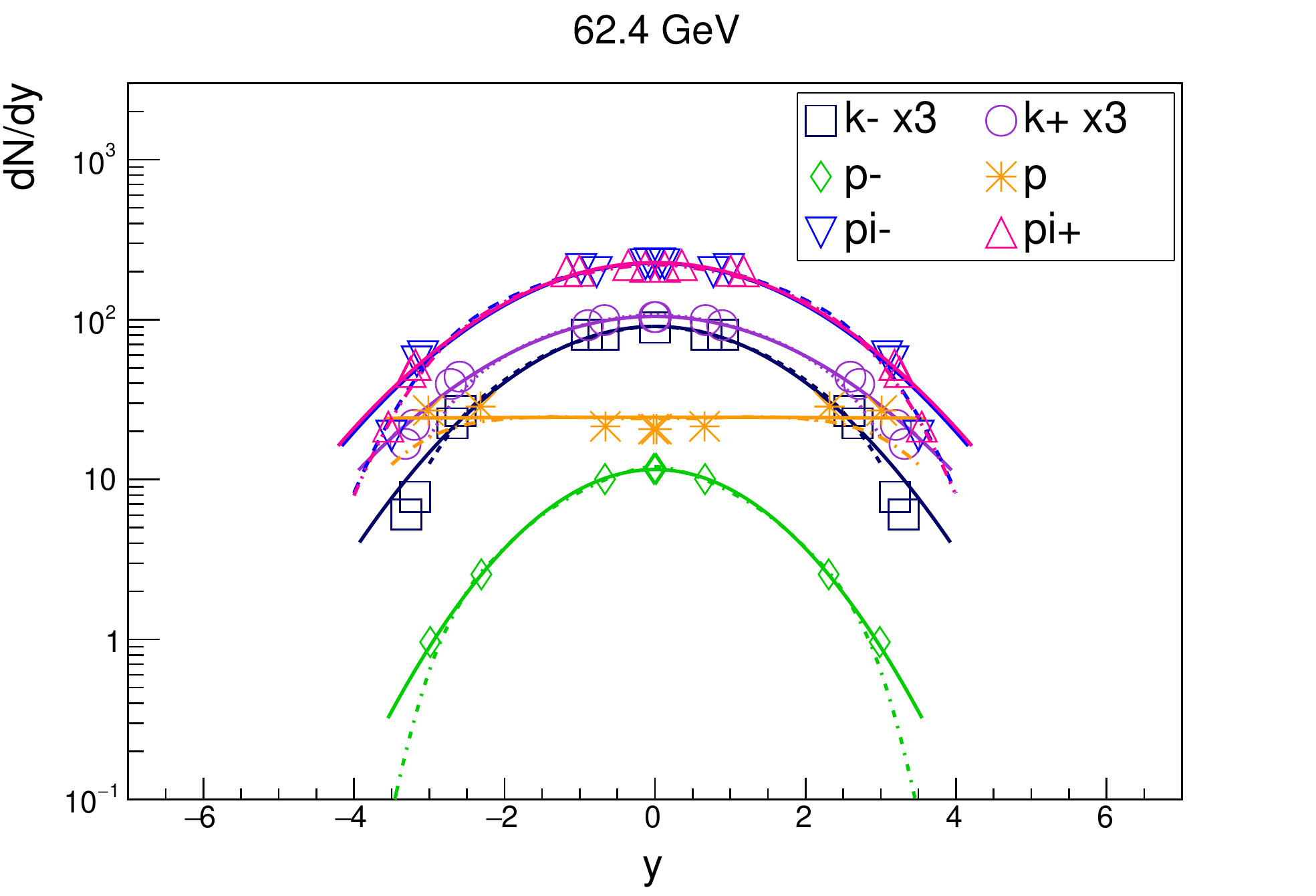}
\caption{The same as in fig.\ref{fig:one} but for Au+Au central collisions at $62.4$ GeV \cite{BRAHMS:2009acd,Arsene:2009aa}. The kaon ranges are multiplied by $3$.}
\label{fig:two}
\end{figure}

In Fig. \ref{fig:two}, the same behavior is found for the particles produced from Au+Au central heavy-ion collisions belong to BRAHMS experiment also but at center of mass energy of $64.2$ GeV \cite{BRAHMS:2009acd,Arsene:2009aa}. Here, we can conclude that all ranges of rapidity distributions for the considered particles are lower than those were at $200$ GeV, as the energy of the collision system decreases. Besides, the EPOS $1.99$ event-generator can successfully reproduce the experimental data of BRAHMS experiment produced from Au+Au central heavy-ion collisions at center of mass energy of $64.2$ GeV very well. Also, the Landau hydrodynamical model shows a good fit with both the experimental measurements and with that estimated from the EPOS $1.99$ event-generator through the obtained minimal chi-square values for the fitting parameters, $N$ and $L$, presented in Tab. (\ref{tab1}).

The rapidity distribution $dN/dy$ of the well-identified hadrons $\pi^{-}$, $\pi^{+}$, $k^{-}$, $k^{+}$, $\bar{p}$, and $p$ measured in BRAHMS experiment (Au+Au collisions), at $\sqrt{s_{NN}}$ = $62.4$ and $200$ GeV, are successfully compared to the EPOS $1.99$ event-generator. The Landau hydrodynamical approach is then fitted to both sets of results. In light of this, we found that the Landau hydrodynamical approach reproduces well the full range of the rapidity distribution for all produced particles, at the considered energies through the minimal values of the chi-square test.

\section{Conclusions}
\label{conc}

We have calculated the rapidity distribution $dN/dy$ for the well-identified hadrons such as $\pi^{-}$, $\pi^{+}$, $k^{-}$, $k^{+}$, $\bar{p}$, and $p$ using the Landau hydrodynamical model. The Landau hydrodynamical approach is then fitted to the experimental measurements from Au+Au central collisions at two different energies, $62.4$ and $200$ GeV, and to that obtained from the corresponding simulations produced from the EPOS $1.99$ event-generator. The excellent agreement between the experimental measurements and simulations gives us a foundation to compare between the two approaches. We found that the Landau hydrodynamical approach successfully reproduces the rapidity distribution across the whole range of rapidity. As a result, we argue that the Landau hydrodynamical approach's statistical assumptions alone may be applied to a wide range of rapidity. Also, the success of the EPOS $1.99$ event-generator to produce the experimental data for the considered particles at the used energies will encourage for further use of it in prediction of the rapidity distribution in regions where the experiment is absent.



\begin {table}[htb]
\caption {The fit parameters obtained from the Landau hydrodynamical approach, Eq. (\ref{equ:three}), for rapidity distribution for the particles $\pi^{-}$, $\pi^{+}$, $k^{-}$, $k^{+}$, $\bar{p}$ and $p$, at the considered energies. Also, the best chi-square values are represented.}
\begin{tabular}{|c|c|c|c|c|}
\hline 
$\sqrt{{S}_{NN}}$ GeV &   particle & $N$ & $L$ & $\chi^{2}$/dof\\ 
\hline 
 & $\pi^{-}$   & $1025.17 \pm 10.76 $ & $3.289 \pm 0.05$ & 0.0082 \\
\cline{2-5}
 &   $\pi^{+}$ & 1044.02 & $3.3588 \pm 0.06$ &0.0078 \\
\cline{2-5}
 $62.4$&    $k^{-}$& $ 119.223 \pm 0.08 $ &  $2.4779 \pm 0.04 $ &0.0092 \\
\cline{2-5}
 &   $k^{+}$ & $163.777 \pm 0.06 $ &  $3.5056 \pm 0.07$ & 0.0095\\
\cline{2-5}
 &  $\bar{p}$  & $2.0202 \pm 0.03$  & $ 0.0026 \pm 0.02$ & 0.092\\
\cline{2-5}
 &  $p$  & $ 4257.49 \pm 0.04 $& $ 4825.62 \pm 40.97 $ & 0.096\\
\cline{2-5}
\hline
 & $\pi^{-}$   & $1809.23 \pm 12.97 $ & $ 5.4785 \pm 0.07 $ & 0.0084\\
 \cline{2-5}
 &   $\pi^{+}$ & $1694.12 \pm 12.65 $ & $5.1688 \pm 0.06 $ & 0.0069\\
 \cline{2-5}
$200$ &   $k^{-}$& $ 226.753 \pm 0.46 $& $4.2672 \pm 0.03$ & 0.0097 \\
 \cline{2-5}
 &   $k^{+}$ & $280.372 \pm 0.03 $& $5.6778 \pm 0.05$  & 0.0099 \\
 \cline{2-5}
 &  $\bar{p}$  & $90.1389 \pm 0.089 $&  $3.7526 \pm 0.05$ &0.074 \\
 \cline{2-5}
 &  $p$  & $231.715 \pm 0.034 $ &  $12.7688 \pm 0.024$ & 0.92\\
 \cline{2-5}
 \hline
 \end{tabular}
\label{tab1}
\end {table}

\end{document}